# Strain Engineering of van Hove Singularity and Coupled Itinerant Ferromagnetism in Quasi-2D Oxide Superlattices

Seung Gyo Jeong[1], Minjae Kim[#,2], Jin Young Oh[1], Youngeun Ham[3], In Hyeok Choi[3], Seong Won Cho[4], Jihyun Kim[1], Huimin Jeong[5], Byungmin Sohn[1], Tuson Park[1], Suyoun Lee[4], Jong Seok Lee[3], Deok-Yong Cho[5, a)], Bongjae Kim[6, a)], and Woo Seok Choi[1, a)]

[1]*Department of Physics, Sungkyunkwan University, Suwon 16419, Republic of Korea*

[#]*Department of Semiconductor Science & Technology, Jeonbuk National University, Jeonju-si 54896, Republic of Korea*

[2]*Korea Institute for Advanced Study, Seoul 02455, Republic of Korea*

[3]*Department of Physics and Photon Science, Gwangju Institute of Science and Technology (GIST), Gwangju 61005, Republic of Korea*

[4]*Center for Semiconductor Technology, Korea Institute of Science and Technology, Seoul 02792, Republic of Korea*

[5]*Department of Physics, Jeonbuk National University, Jeonju 54896, Republic of Korea*

[6]*Department of Physics, Kyungpook National University, Daegu 41566, Republic of Korea*

[a)] Author to whom correspondence should be addressed: zax@jbnu.ac.kr, bongjae@knu.ac.kr, choiws@skku.edu

**ABSTRACT**

Engineering van Hove singularities (vHss) near the Fermi level, if feasible, offers a powerful route to control exotic quantum phases in electronic and magnetic behaviors. However, conventional approaches, which rely primarily on chemical and electrical doping, focus mainly on local electrical or optical measurements, limiting their applicability to coupled functionalities. In this study, a vHs-induced insulator-metal transition coupled with a ferromagnetic phase transition was empirically achieved in atomically designed quasi-2D $SrRuO_3$ (SRO) superlattices via epitaxial strain engineering, which has not been observed in conventional 3D SRO systems. Theoretical calculations revealed that epitaxial strain effectively modulates the strength and energy positions of vHs of specific Ru orbitals, driving correlated phase transitions in the electronic and magnetic ground states. X-ray absorption spectroscopy confirmed the anisotropic electronic structure of quasi-2D SRO modulated by epitaxial strain. Magneto-optic Kerr effect and electrical transport measurements demonstrated modulated magnetic and electronic phases. Furthermore, magneto-electrical measurements detected significant anomalous Hall effect signals and ferromagnetic magnetoresistance, indicating the presence of magnetically coupled charge carriers in the 2D metallic regime. This study establishes strain engineering as a promising platform for tuning vHss and resultant itinerant ferromagnetism of low-dimensional correlated quantum systems.

**I. INTRODUCTION**

Correlated low-dimensional metals frequently host van Hove singularities (vHss), arising from the saddle points in their electronic band structure, leading to local divergence in the density of states (DOS).[1,2] Unlike conventional 3D systems, the diverging 2D vHs, if tunable, would provide substantial modulation in a variety of physical and chemical properties. When the energy position of the vHs is near the Fermi level ($E_F$), electronic instabilities become active, giving rise to or manipulating correlated quantum phenomena. These include unconventional magnetism, superconductivity, charge density wave, spin density wave, nontrivial topology, quantum and topological Hall effects, and metal-insulator transitions.[3-8] The intricate interplay among charge, lattice, spin, and orbital degrees of freedom, combined with the dimensionality of the system, offers effective control over these emergent properties by tuning the vHs.[9-12] Several strategies have been developed to access vHss experimentally, including chemical doping[13-16] or electric-field gating[17,18] in different materials and stacking angle control in bilayer graphene.[19-21] While the electronic band structures containing vHss have often been characterized using scanning tunneling microscopy/spectroscopy or angle-resolved photoemission spectroscopy,[13-21] coupled electromagnetic phase transitions with functionalities through vHs tuning remain to be discovered in most materials.

Epitaxial strain engineering of quasi-2D SrRuO$_3$ (SRO) provides a critical control knob for vHs-induced phase transitions in both magnetic and electronic ground states. Epitaxial strain has been successfully used to modulate electronic and magnetic properties in ultrathin transition metal oxides.[22-25] Especially, The low-energy electronic structure of quasi-2D SRO is governed by $t_{2g}$ electrons in the Ru (4$d$) shell.[26-33] Specifically, the in-plane bonding of $d_{xy}$ orbitals forms a 2D vHs near $E_F$, as schematically illustrated in Fig. 1(a), whereas the truncated out-of-plane bonding of $d_{yz/zx}$ orbitals exhibits quasi-1D electronic structures, each hosting a 1D singularity near the band edges.[26-33] Theoretical calculations suggest that epitaxial strain can effectively modulate the crystal field splitting between the $d_{xy}$ and $d_{yz/zx}$ orbitals.[5,34] This modulation alters the energy position of the vHs of the $d_{xy}$ band near $E_F$, consequently affecting the orbital occupancies. Such changes can drive the system toward or away from the Mott-insulating phase, which is characterized by half-filled $d_{yz/zx}$ and fully occupied $d_{xy}$ orbitals.[35] Furthermore, changes in the low-energy electronic structure can disturb the delicate balance between competing magnetic instabilities.[5,34,36] Specifically, when the vHs energy position approaches $E_F$, Stoner-type itinerant ferromagnetism is predicted to emerge, prevailing over the Mott-insulating antiferromagnetic phase.[5]

In this study, we experimentally realized a 2D vHs-induced metal-insulator transition coupled with a ferromagnetic transition in an atomically designed quasi-2D SRO through epitaxial strain engineering. By combining density functional theory with dynamical mean-field theory (DFT+DMFT) calculations, we found that epitaxial strain effectively modulates the vHs, facilitating correlated phase transitions in the electronic and magnetic ground states of quasi-2D SRO. Experimentally, X-ray absorption spectroscopy confirmed the modulation of the anisotropic electronic structure of quasi-2D SRO induced by epitaxial strain. The resulting magnetic and electronic phase transitions were verified through the magneto-optic Kerr effect and electrical transport measurements. Electrical transport measurements under a magnetic field further revealed a pronounced anomalous Hall effect and ferromagnetic magnetoresistance in the compressively strained quasi-2D SRO, highlighting the strong influence of vHs in determining the electronic and magnetic phases. The results highlight the facile engineering of low-dimensional ferromagnetic metallic systems via strain-controlled manipulation of the 2D vHs.

## II. RESULTS AND DISCUSSION

To achieve an artificial quasi-2D SRO system, we designed epitaxial superlattices consisting of alternating atomic monolayers of perovskite SRO and $SrTiO_3$ (STO), that is, [(SRO)$_1$|(STO)$_1$]. Bulk SRO is an itinerant ferromagnet with a transition temperature ($T_c$) of ~150 K.[37] However, the ferromagnetic behavior of atomically thin SRO becomes weaker, accompanied by a metal-insulator transition below a few atomic unit cells (single perovskite unit cell (u.c.) = ~0.4 nm).[37] Epitaxial superlattices enable the confinement of SRO within a single unit cell, effectively hosting 2D vHs features while exhibiting observable magnetic and electrical signals owing to interlayer exchange interactions.[38-40] The presence of insulating STO monolayers on both sides of monolayer SRO spatially confines the electronic states along the out-of-plane direction, leading to an effective 2D confinement. On the other hand, finite interlayer coupling between adjacent SRO monolayers across the STO monolayer spacers leads to robust magnetic behavior in the superlattice. Hence, we introduce the term "quasi-2D" to define the dimensional nature of the [(SRO)$_1$/(STO)$_1$] superlattices. Fig. 1(b) schematically demonstrates tunable vHs with accessible electromagnetic phases of quasi-2D SRO subject to in-plane epitaxial strain. Note that the direction of the in-plane strain is reversed between the upper and lower panels. The vHs of the $d_{xy}$ orbitals in quasi-2D SRO superlattices is located below $E_F$,[5,34] which is distinct from its bulk counterpart Sr$_2$RuO$_4$.[28,41] Moreover, because of the stronger electronic correlation arising from confinement, the quasi-2D SRO superlattice tends to exhibit stronger magnetic behavior,[34] making the system even more susceptible to external perturbations.

DFT+DMFT calculations provided key insights into the epitaxial-strain-tunable vHs with coupled magnetism in quasi-2D SRO ([(SRO)$_1$|(STO)$_1$]), as shown in Figs. 1(c)-1(h). Under compressive strain, the in-plane Ru-O bond length decreases relative to the out-of-plane Ru-O bond length (Fig. S1), thereby increasing the energy level of the $d_{xy}$ orbitals relative to that of the $d_{yz/zx}$ orbitals [Fig. 1(c)] and modifying the occupancy of the orbitals [Fig. 1(d)]. This shift brings the $d_{xy}$ orbital-induced vHs closer to $E_F$ [Fig. 1(b)], increasing the DOS at $E_F$ [Figs. 1(e) and 1(f)] (Fig. S2) and enhancing the ferromagnetic instability via the Stoner criterion. The calculated ferromagnetic susceptibility ($\chi$) under compressive strain [Fig. 1(g)] further confirms this orbital-selective behavior, where the significant contribution from $d_{xy}$ orbitals induces ferromagnetism through strain-induced vHs tuning. On the other hand, tensile strain increases the crystal field splitting energy ($h_{CF}$) between the $d_{xy}$ and $d_{yz/zx}$ orbitals, as shown in Fig. 1(c). Although this also increases the total DOS at $E_F$ and enhances total $\chi$ [Figs. 1(f) and 1(h)], the underlying physics is fundamentally different from that of the compressive strain case. The tensile strain increases the DOS of $d_{yz/zx}$ but decreases that of $d_{xy}$ [Fig. 1(e)], resulting in the enhancement of total DOS at $E_F$ (Fig. 1f). More importantly, the tensile strain progressively drives the $d_{xy}$ orbital toward full occupancy, whereas the $d_{yz/zx}$ orbitals approach a half-filled regime, as shown in Fig. 1(d) (and also see Fig. S3, supplementary material). This orbital-selective occupation under tensile strain pushes the [(SRO)$_1$|(STO)$_1$] system toward a Mott-insulating state. In this regime, a Néel-type antiferromagnetic ground state is predicted from the superexchange between the localized spins, S = 1, of the $d_{yz/xz}$ orbitals, similar to that of Ca$_2$RuO$_4$ at low temperatures.[42-44] Unlike the case for compressive strain, where the $d_{xy}$ orbitals predominantly contribute to χ [Fig. 1(g)], tensile strain results in a comparable contribution from the $d_{yz/zx}$ orbitals, which exhibit pronounced competing antiferromagnetic interactions characteristic of the Mott-insulating state with half-filled $d_{yz/zx}$ orbitals.

To validate the theoretically proposed scenario, we synthesized quasi-2D SRO superlattices under various in-plane strain states (from −1.60% to +1.21%) using pulsed laser epitaxy. Fig. 2(a) shows the X-ray diffraction (XRD) θ-2θ scan results of [(SRO)$_1$|(STO)$_1$]$_{10}$ superlattice epitaxially grown on STO (001) substrate. At a small $q_z$, where low grazing angles allow X-ray specular reflections, the intensity oscillates with the wave vector $q_z$, forming Kiessig fringes. It starts to reveal the lowest superlattice peak satisfying the Bragg condition at $q_z$ = 0.8 Å$^{-1}$, corresponding to the supercell structure of quasi-2D SRO composed of alternating SRO and STO monolayers. With increasing $q_z$, we observed the (001) and (002) Bragg diffraction peaks of SRO and STO at $q_z$ = 1.6 and 3.2 Å$^{-1}$, respectively, and a half-order superlattice diffraction peak at $q_z$ = 2.4 Å$^{-1}$. Multiple Laue oscillations were observed between the Bragg diffractions. The presence of Kiessig fringes and Laue oscillations consistently indicates atomically sharp interfaces between SRO and STO and the high crystallinity of the artificial quasi-2D SRO superlattices. Fig. 2(b) shows the XRD reciprocal space mapping (RSM) results for the strain-engineered quasi-2D SRO superlattices grown on various perovskite oxide substrates: NdGaO$_3$ (NGO), STO, DyScO$_3$ (DSO), TbScO$_3$ (TSO), and GdScO$_3$ (GSO). The RSMs were measured around the (103) Bragg diffraction peaks of each substrate in (pseudo-)cubic notation. These results exhibit identical in-plane wave vectors ($q_x$) of the substrate (intense, red) and film peaks (broad, green), indicating fully strained states of the quasi-2D SRO superlattices. Systematic modulation of the out-of-plane lattice depending on the epitaxial strain was also confirmed in XRD θ-2θ scans for all the superlattices, as shown in Fig. S4, supplementary material. The SRO/STO superlattice grown on LaAlO$_3$ exhibited in-plane strain relaxation because of a large lattice mismatch (Fig. S5, supplementary material). Hence, we focused on the fully strained quasi-2D SRO superlattices to isolate the strain relaxation effect. Assuming the lattice constant of quasi-2D SRO is 3.916 Å, which is the average value of bulk SRO (3.926 Å) and STO (3.905 Å), it was estimated that the in-plane strain states of quasi-2D SRO superlattices on NGO, STO, DSO, TSO, and GSO substrates are −1.60%, −0.27%, +0.73%, +1.06%, and +1.21%, respectively. Negative and positive signs indicate compressive strain and tensile strain, respectively.

First, the anisotropy variations in the Ru 4$d$ orbitals were examined to verify the strain-engineered electronic structures in the quasi-2D SRO. Figs. 2(c)-2(e) show X-ray absorption spectroscopy (XAS) and X-ray linear dichroism (XLD) results at the Ru $L_3$-edge for the superlattices. Here, we measured three representative superlattices with the largest strain differences: NGO (−1.60%,), STO (−0.27%), and GSO (+1.21%), demonstrating clear anisotropy evolution of the Ru 4$d$ local electronic character. By performing XAS using a rotatable sample holder (see the Experimental Section for further details), we obtained XAS spectra with two different light polarizations; one along the in-plane direction ($E//xy$) and the other mostly along the out-of-plane direction ($E//z$) with respect to the pseudocubic crystal axes. The isotropic term of XAS peak intensities, ($I_{E//z} + 2I_{E//xy}$)/3, at the Ru $L_3$-edge reflect the overall Ru $2p_{3/2}$ to 4$d$ transition near a photon energy of 2843 eV.[45,46] All spectra show negligible energy shifts with similar bandwidths of the excitation peaks, indicating that the Ru ions predominantly maintain a +4 oxidation state.[45-47] For Ru 4$d$ orbitals, the transitions to $d_{yz/zx}$ and $d_{z2}$ orbitals contribute to $I_{E//z}$, whereas the transitions to $d_{xy}$, $d_{yz/zx}$, and $d_{x2-y2}$ orbitals contribute to $I_{E//xy}$. XLD is defined as the difference between these intensities: XLD = $I_{E//xy} - I_{E//z}$. Although the XLD signals are relatively weak compared to the XAS intensities, they systematically reflect the energy ordering of each orbital state. The subsequent peak and dip (or dip and peak) structures observed in the XLD signals at a lower energy (~2840 eV) and higher energy (~2846 eV), respectively, reveal the relevant nature of the unoccupied Ru 4$d$ $e_g$ orbitals, indirectly reflecting the anisotropic electronic structures of the quasi-2D SRO. In particular, the peaks are primarily related to $d_{x2-y2}$ orbitals, whereas the dips are mainly associated with $d_{z2}$ orbitals.

It is clear that the $d_{z^2}$ orbital has a lower (higher) energy than the $d_{x^2-y^2}$ orbitals in the case of the NGO substrate (STO and GSO substrates). The strong substrate dependence indicates that epitaxial strain serves as an effective control knob for the electronic structure of quasi-2D SRO superlattices. The simulation results shown in Fig. S6, supplementary material, appear consistent with the experimental XLD results. Here, the substrate dependence was mimicked by adjusting the Ru-O orbital hybridization strengths only from the parameter set for the quasi-2D SRO on STO substrate. The hybridization strength of the Ru $d_{z^2}$-apical O $2p$ was preferentially reduced by half to account for the loosening of apical bonds resulting from the compressive strain. For the GSO substrate, all Ru $4d$-O $2p$ hybridization strengths were reduced to 75%, reflecting overall bond weakening caused by tensile strain, which is consistent trend with covalency reduction of a Mott-insulating state in the DFT + DMFT calculations (Fig. S7, supplementary material). The substrate-dependent hybridization evolution was further confirmed by the O $K$-edge XAS spectra shown in Figs. 2(f)-2(h), where the intensities of the first peaks at ~530.5 eV were approximately proportional to the Ru $4d$-O $2p$ hybridization strengths. The difference between $I_{E//xy}$ and $I_{E//z}$ was most pronounced in the quasi-2D SRO on STO substrate, whereas it decreased in both superlattices on the NGO and GSO substrates, which is consistent with the Ru $L_3$-edge XLD results.

Fig. 3 shows the strongly coupled electromagnetic and optical properties of the quasi-2D SRO originating from the epitaxial-strain-tuned vHs. We first show the magneto-optic Kerr effect (MOKE) measurements demonstrating compressive strain-induced ferromagnetism, as shown in Figs. 3(a)-3(c). We used polar MOKE geometry with a laser energy of 1.6 eV (below the STO band gap of 3.2 eV) to selectively detect the out-of-plane magnetization of the quasi-2D SRO layer. Fig. 3a shows the temperature ($T$)-dependent Kerr angle ($\theta_K$) for quasi-2D SRO superlattices obtained through a field-cooled-warming process with an out-of-plane magnetic field ($H$) of 0.1 T. The $\theta_K$ ($T$) curves of superlattices on NGO, STO, and DSO substrates exhibit a distinct signature of the ferromagnetic order with a variation of transition temperature ($T_c$), while the superlattice on GSO substrate shows a negligible signal within the measured temperature range down to 25 K. We confirmed the ferromagnetic behavior of the quasi-2D SRO superlattice on STO substrate using a superconducting quantum interference device (SQUID) magnetometer, as shown in Fig. S7, supplementary material. Fig. S8 shows ferromagnetic hysteresis loops with a coercive field of ~0.2 T, comparable to that of bulk-like 60-nm-thick SRO films on STO.[45] The reduced saturation and remnant magnetization compared to single SRO films are likely due to the dimensional confinement effects.[39] Note that NGO and GSO substrates are magnetic, and hence SQUID measurements on the superlattices on those substrates were not reliable. The strain-dependent $T_c$ and $\theta_K$ ($T$ = 25 K) of quasi-2D SRO superlattices are summarized in Figs. 3(b) and 3(c), respectively. Both $T_c$ and $\theta_K$ ($T$ = 25 K) increase with compressive strain, suggesting that compressive strain induces 2D ferromagnetism in SRO monolayers. Note that $\theta_K$ is generally proportional to the magnetization in ferromagnets.[48] Hence, the larger $\theta_K$ ($T$ = 25 K) for superlattices with compressive strain indicates strain-controlled vHs and a resultant larger $\chi$, consistent with the theoretical calculations discussed in Fig. 1.

Strong coupling between the electronic and magnetic degrees of freedom was observed in the strain-induced ferromagnetic phase, which coincided with the 2D metallic phase. Figs. 3(d)-3(f) show the $T$-dependent electrical resistivity ($\rho$ ($T$)) of strain-controlled quasi-2D SRO superlattices. The Mott–Ioffe–Regel limit, determined by $k_F \lambda \approx 1$ - $2\pi$,[49] where $k_F$ and $\lambda$ are the Fermi wave vector and mean free path of the electrons in SRO, respectively, indicates the boundary of $\rho$ values for the Mott transition. The tensile-strained quasi-2D SRO superlattices on DSO, TSO, and GSO substrates exhibited insulating $\rho$ ($T$) behavior, with $\rho$ values exceeding the Mott–Ioffe–Regel limit in the low-$T$ region. In contrast, the compressively strained quasi-2D SRO superlattices on NGO and STO substrates exhibited semiconducting behavior, with $\rho$ values near the lower boundary of the Mott–Ioffe–Regel limit. The electrical transport results suggest the presence of a 2D metallic phase in the compressively strained quasi-2D SRO films, even though they exhibited a minute upturn in $\rho$ ($T$) at the low-$T$ region. This may result from the discontinuity of the monolayer metallic film caused by the step-terrace surface structure of the substrate and/or potential disorder in the films. Fig. S9 shows step-terrace structures of the quasi-2D SRO superlattices grown on STO and GSO substrates, suggesting that step-terrace-induced discontinuities may act as a potential source of disorder, resulting in an increase of electrical resistivity at low temperature. Nevertheless, the structural consistency cannot account for the metallic and insulating behavior of the superlattices grown on STO and GSO substrates, respectively. We summarize the strain-dependent $\rho$ ($T$ = 10 K) in Fig. 3(e), demonstrating a strain-induced metal-insulator transition in quasi-2D SRO superlattices. The thermal activation energy gap ($E_a$) of the insulating phases was estimated using the Arrhenius equations,[47] as summarized in Fig. 3(f). Whereas the tensile-strained superlattices exhibited a large $E_a$ of ~10 meV, indicating Mott-insulating phases, the compressively strained superlattices showed small gaps that emerge only at low $T$s, because of potential disorder. Such strain-induced ferromagnetic and metal-insulator transitions have not been reported in conventional 3D SRO,[50-52] highlighting a unique advantage of the quasi-2D SRO systems in this study. Here, it was discovered experimentally that the distinctive vHs properties of quasi-2D SRO induce coupled electronic and magnetic phase transitions through epitaxial strain engineering.

The ferromagnetic metallic phases of the compressively strained quasi-2D SRO further revealed an anomalous Hall effect (AHE), as shown in Figs. 3(g)-3(j). Figs. 3(g), 3(h), and 3(i) display the $H$-dependent Hall resistivity ($\rho_{xy}$ ($H$)) of superlattices on NGO, STO, and GSO substrates, respectively, at selected $T$ values. The extended $T$ dependencies are shown in Figs. S10-S12, supplementary material. For the compressively strained quasi-2D SRO superlattices (on the NGO and STO substrates in Figs. 3(g) and 3(h), respectively), a hysteretic AHE was observed. The AHE contribution ($\rho_{AHE}$) was extracted by subtracting the ordinary Hall effect contribution ($\rho_{OHE}$) using linear fitting above the coercive field from $\rho_{xy}$ (Fig. S13), for the superlattice on the STO substrate at 25 K. Fig. 3(j) summarizes the extended $T$ dependence of $\rho_{AHE}$ values at $H$ = 4.5 T for the superlattices on NGO and STO substrates. As $T$ decreases, the superlattice on the NGO substrate exhibits a much steeper $T$-dependent $\rho_{AHE}$ slope compared to that on the STO substrate. In the low-$T$ regions (below 70 K), the $\rho_{AHE}$ values become larger for the superlattice on the NGO substrate, consistent with the MOKE results shown in Fig. 3(a). This result reveals a strong coupling between the ferromagnetic order and spin-polarized itinerant charge carriers in quasi-2D SRO. Given that both intrinsic (Berry curvature) and extrinsic (skew scattering) mechanisms are known in SRO,[53-55] the AHE in our superlattices likely arises from their combined contributions. Conversely, the tensile-strained film on the GSO substrate in Fig. 3(i) shows only a linear slope of $\rho_{OHE}$, with no observable AHE down to 45 K. Hall effect measurements at lower $T$ could not be obtained because the superlattice on the GSO substrate was highly insulating (Fig. S12, supplementary material).

The $H$-dependent magnetoresistance of the quasi-2D SRO reveals that ferromagnetic hysteresis appears only in the compressively strained superlattices [Figs. 3(k)-3(n)], consistent with the Hall effect results. Magnetoresistance measurements enabled us to investigate the magneto-transport properties of insulating samples down to a lower $T$ of 15 K. Figs. 3(k) and 3(l) show the characteristic butterfly-shaped ferromagnetic hysteresis behavior at 15 K for the superlattices on the NGO and STO substrates, respectively. In contrast, Fig. 3(m) shows a parabolic shape without hysteresis for the superlattice on the GSO substrate. This indicates the nonferromagnetic behavior of the tensile-strained quasi-2D SRO superlattices, which aligns with the MOKE and Hall measurements. Fig. 3(n) summarizes the magnetoresistance value at 4.5 T as a function of $T$. Negative magnetoresistance values in the three differently strained films exhibited an anomaly in the low-$T$ region, which have been observed in both ferromagnetic and nonferromagnetic SRO monolayers.[56,57] The absence of ferromagnetism in insulating tensile-strained quasi-2D SRO superlattices can be associated with the reduction in the DOS of $d_{xy}$ orbitals at the $E_F$, as suggested by the Stoner model. Thus, the quasi-2D ferromagnetic metallic state can be realized only under compressive strain, achieved by shifting the position of the 2D vHs peak, as shown in Fig. 1.

### III. CONCLUSION

In summary, an epitaxial-strain-engineered 2D vHs and correlated phase transitions between ferromagnetic metallic and nonferromagnetic insulating states in quasi-2D SRO superlattices were demonstrated. Guided by DFT+DMFT calculations, we employed XAS, MOKE, and magneto-electrical transport measurements to confirm the modulation of anisotropic orbital occupations, metal-insulator transitions, and coupled ferromagnetic transitions in quasi-2D SRO superlattices by controlling the epitaxial strain. We emphasize such phenomena have not been observed in conventional 3D SRO systems.[50-52] Compressively strained quasi-2D ferromagnetic metallic phases exhibited an AHE with hysteresis that could be further tuned via epitaxial strain. Theoretical calculations attribute these correlated functional phenomena to the strain-modulated vHs below the Fermi level in quasi-2D SRO. This behavior is distinct from that of the natural crystal structure of the layered perovskite $Sr_2RuO_4$, where a vHs above the Fermi level induces controllable superconductivity through uniaxial pressure.[58] Considering the stronger locality and magnetic instability of the superlattice system in this study, superconductivity with even higher critical temperatures than its bulk counterpart may be envisaged. This further suggests that the energy position of the 2D vHs relative to the $E_F$ determines various functionalities, highlighting the potential for controlling the 2D vHs. Although direct momentum-resolved observation of strain-modulated vHs is hindered by the insulating STO topmost layers, our results provide evidence for its role in the observed electronic and magnetic transitions. Building on previous spectroscopic studies of SRO thin films[35] and $Sr_2RuO_4$[27] showing strain-tunable vHs near the Fermi level, we demonstrate how such vHs couple to ferromagnetism, metal-insulator transitions, and AHE in quasi-2D SRO superlattices. This epitaxial strain approach paves a practical pathway for realizing correlated functionalities in systems with 2D vHs through the artificial design of synthetic crystals.

## IV. METHODOLOGY

### A. Pulsed laser epitaxy and structural characterizations

Atomically designed epitaxial [(SRO)$_1$|(STO)$_1$] superlattices with 10 repetitions were synthesized on various single-crystal perovskite oxide substrates using pulsed laser epitaxy.[59] Both the SRO and STO layers were deposited at 750 °C under 100 mTorr of oxygen partial pressure from stoichiometric ceramic targets using a KrF excimer laser (248 nm; IPEX-868, Light Machinery). A laser fluence of 1.5 Jcm$^{-2}$ and a repetition rate of 5 Hz were employed. The superlattice thicknesses were determined by X-ray reflectivity, whereas the crystalline structures and strain states of the epitaxial thin films were confirmed through XRD $\theta$–$2\theta$ and reciprocal space map scans using high-resolution XRD systems (PANalytical X'Pert and a Rigaku Smartlab XRD). The Bragg peaks of the superlattice structures were measured using synchrotron XRD at the 5D beamline of the Pohang Light Source at room temperature.

### B. Theoretical calculations

For the crystal structure of the [(SRO)$_1$|(STO)$_1$] superlattice, we adopted the optimized structures obtained within the DFT framework using the gradient generalized approximation (GGA) exchange-correlation functional[34], and verified that the structures using the local density approximation (LDA) functional showed negligible differences. The biaxial strain is referenced based on the averaged in-plane lattice constant of bulk SRO and STO, 3.916 Å. The octahedral rotation ($\sqrt{2}\times\sqrt{2}$) is allowed ($a^0a^0c^-$ Glazer tilting of octahedra) in the tetragonal unit cell. Further details can be found elsewhere.[34] The present computational setup for the DFT + DMFT framework is the same as that in the literature.[5] Electronic structure calculations in the DFT + DMFT framework[60,61] using the $t_{2g}$-only projector method were performed using the full potential implementation in the TRIQS library.[62,63] The DFT part of the computations in the local density approximation was performed employing the WIEN2k package.[64] We used 16 × 16 × 11 $k$-point mesh for the Brillouin zone integration. Wannier-like $t_{2g}$ orbitals were constructed from the Kohn-Sham bands, which included six $t_{2g}$ bands and the seven uppermost O 2$p$ bands. We used the full rotationally invariant Kanamori interaction with $U$ = 2.3 eV and $J$ = 0.4 eV. This setup successfully described the physical properties of the ruthenates in previous studies.[5,31,44,65-68] The

quantum impurity problem in DMFT was solved using the continuous-time hybridization-expansion quantum Monte Carlo impurity solver as implemented in the TRIQS library.[69,70] The temperature was set to 35 K in the DFT + DMFT calculations. We performed charge self-consistent DFT + DMFT computation for the paramagnetic phase with the convergence criteria of the total energy of 0.002 Ry/formula unit. For the computation of the ferromagnetic susceptibility ($\chi$), the converged charge densities from the paramagnetic calculations were fixed. We allowed spin symmetry breaking in the local Green's function and the self-energy of the $t_{2g}$ orbital according to the ferromagnetic order and performed DMFT iterations until the electronic structure convergence in the presence of the ferromagnetic field. The ferromagnetic field for the computation of $\chi$ was set at 5 meV (approximately 86 T).

**C. X-ray absorption spectroscopy**

The Ru $L_3$-edge and O $K$-edge XAS spectra were obtained at the 6A beamline of the Pohang Light Source by measuring the total electron yield, which is sensitive to the film surface. Measurements were performed at room temperature. To obtain the X-ray polarization ($E$) dependence and XLD signal, the samples were positioned relative to the direction of the incident X-ray either at 0° (surface normal, $E//xy$) or rotated to 70° (a combination of $E//xy \times \cos^2 70°$ and $E//z \times \sin^2 70°$), where $xy$ and $z$ denote the crystal axes. XLD simulations were conducted using CTM4XAS.[71] The detailed simulation parameters are provided in the supplementary materials.

**D. Electrical transport and magnetization measurements**

The electrical transport properties of the superlattices, including longitudinal resistivity ($\rho$) and Hall resistivity ($\rho_{xy}$) were measured using the van der Pauw method. A current source (K2612, Keithley) and a nanovoltmeter (K2182, Keithley) were used inside a closed-cycle cryostat (CMag Vari.-9, Cryomagnetics) equipped with a 9 T superconducting magnet. To extract the anomalous $\rho_{xy}$ ($\rho_{AHE}$), the contribution of the ordinary Hall effect in $\rho_{xy}$ was subtracted based on a linear fitting of the data above the coercive field. Magnetization measurements were performed using a SQUID magnetometer in a magnetic property measurement system (MPMS, Quantum Design). Temperature-dependent magnetization, $M(T)$, was measured using the field-cooled method with a 0.01 T magnetic field ($H$) applied along the out-of-plane direction, and $H$-dependent magnetization, $M(H)$, was recorded with out-of-plane $H$-fields. We conducted MOKE measurements using a pulsed laser (Vision-S, Coherent, 80 MHz repetition rate) with a center wavelength of 800 nm. To minimize the steady-state heating effect, we used the unfocused laser beam with a fluence of approximately 7 nJ cm$^{-2}$. The beam was modulated by a mechanical chopper (700 Hz) and a photoelastic modulator (Hinds, 100 kHz). The Kerr angle signal was obtained using a digital lock-in amplifier (Zurich). We performed $T$-dependent MOKE measurements using a closed-cycle helium cryostat (Montana) with a field-cooled-warming process while an $H$ of 0.1 T was applied along the out-of-plane direction.

**SUPPLEMENTARY MATERIAL**

See the supplementary material for in-plane strain dependent Ru-O distance of the quasi-2D SRO superlattices for the DFT + DMFT computation, calculated quasiparticle residues and occupancy of orbitals as a function of in-plane strain, XRD $\theta$-$2\theta$ scan results of the quasi-2D SRO superlattices grown on various substrates, XRD reciprocal space mapping result of the SRO/STO superlattice grown on LaAlO$_3$ (001) substrate, Linear dichroisms, obtained from experimental and simulated results, in-plane strain and electronic phase dependent Ru($t_{2g}$) occupancy, magnetization measurements as a function of $T$ and $H$-field, Raw data of $H$-field-dependent sheet resistance and Hall resistance for quasi-2D SRO superlattices grown on NGO, STO, and GSO, extracted $\rho_{AHE}$ signal after subtracting $\rho_{OHE}$ of quasi-2D superlattice on STO substrate.

**ACKNOWLEDGMENTS**

This work was supported by the National Research Foundation of Korea (NRF) grant funded by the Korea government (MSIT) (Nos., 2021R1A2C2011340, RS-2023-00220471, RS-2023-00281671, and RS-2022-NR068223). M.K. was supported by Korea Institute for Advanced Study (KIAS) individual Grants (No. CG083502). The DFT + DMFT calculation was supported by the Center for Advanced

Computation at KIAS. B.K. acknowledges support from NRF grants No. 2021R1C1C1007017, No. 2021R1A4A1031920, and No. RS-2022-NR068223 and the KISTI Supercomputing Center (Project No. KSC-2022- CRE-0465). S.L. was supported by the National Research Council of Science & Technology (NST) grant by the Korea government (MSIT) (No. GTL24041-000).

**AUTHOR DECLARATIONS**

**Conflict of Interest**

The authors declare that they have no known competing financial interests or personal relationships that could have appeared to influence the work reported in this paper.

**Author Contributions**

Seung Gyo Jeong, Minjae Kim, and Jin Young Oh contributed equally to this work.

Seung Gyo Jeong: Conceptualization (lead); Methodology (lead); Investigation (lead); Visualization (lead); Writing – original draft (lead); Writing – review & editing (lead). Minjae Kim: Conceptualization (equal); Methodology (equal); Investigation (equal); Writing – original draft (equal); Writing – review & editing (equal). Jin Young Oh: Conceptualization (equal); Methodology (equal); Writing – original draft (lead); Writing – review & editing (equal). Youngeun Ham: Methodology (supporting); Writing – review & editing (supporting). In Hyeok Choi: Methodology (supporting); Writing – review & editing (supporting). Seong Won Cho: Methodology (supporting); Writing – review & editing (supporting). Jihyun Kim: Methodology (supporting); Writing – review & editing (supporting). Huimin Jeong: Methodology (supporting); Writing – review & editing (supporting). Byungmin Sohn: Methodology (supporting); Writing – review & editing (supporting). Tuson Park: Methodology (supporting); Writing – review & editing (supporting). Suyoun Lee: Methodology (supporting); Writing – review & editing (supporting). Jong Seok Lee: Methodology (supporting); Writing – review & editing (supporting). Deok-Yong Cho: Conceptualization (equal); Methodology (equal); Investigation (equal); Funding acquisition (lead); Project administration (lead); Supervision (lead); Writing – original draft (lead); Writing – review & editing (equal). Bongjae Kim: Conceptualization (equal); Methodology (equal); Investigation (equal); Funding acquisition (lead); Project administration (lead); Supervision (lead);

Writing – original draft (lead); Writing – review & editing (equal). Woo Seok Choi: Conceptualization (equal); Methodology (equal); Investigation (equal); Visualization (supporting); Funding acquisition (lead); Project administration (lead); Supervision (lead); Writing – original draft (lead); Writing – review & editing (lead).

## DATA AVAILABILITY

All data supporting the findings and conclusions of this study are provided in the main text and supplementary materials.

**Figures**

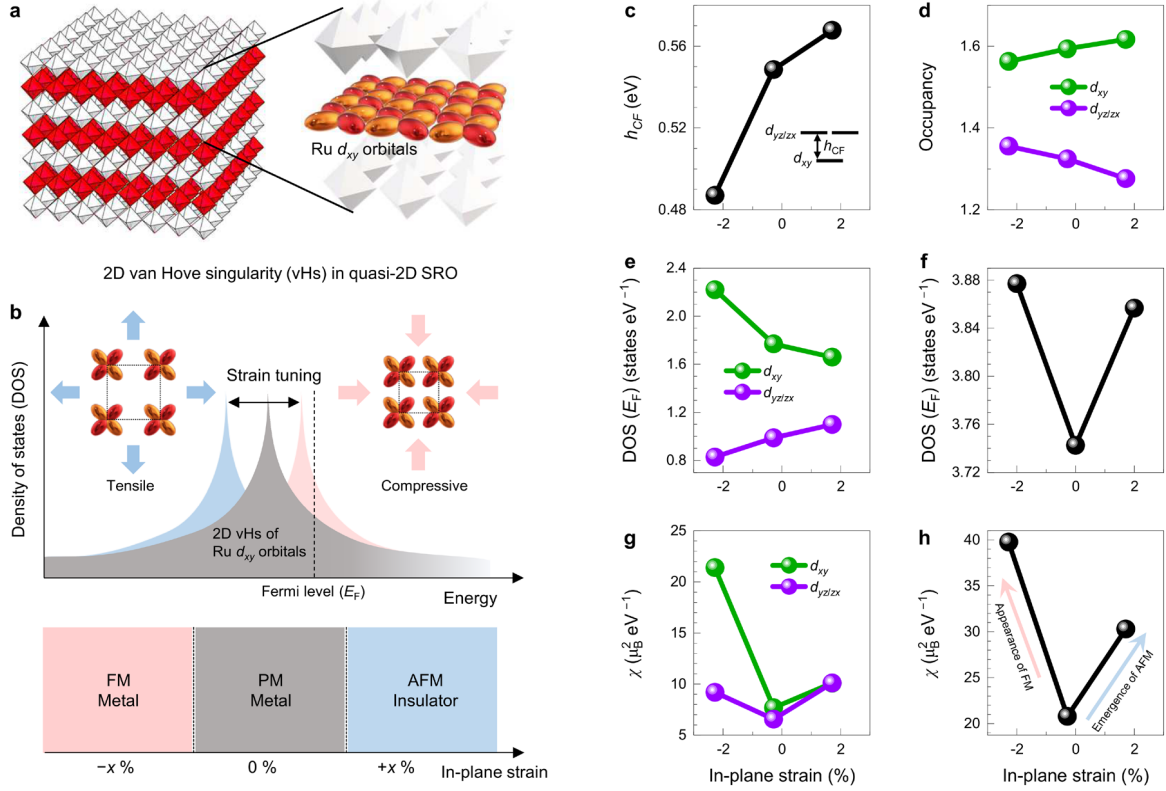

**Fig. 1. Concepts and theoretical investigations of epitaxial strain controlled 2D vHs and coupled electromagnetic phases in quasi-2D SRO systems.** (a) Schematic illustration of the 2D Ru $d_{xy}$ orbitals in quasi-2D SRO, i.e., the [(SRO)$_1$|(STO)$_1$] superlattice. Here, STO is SrTiO$_3$. The white and red octahedra represent TiO$_6$ and RuO$_6$, respectively. (b) Sketch of epitaxial-strain-tuned 2D vHs of quasi-2D SRO below the Fermi level ($E_F$) from the density of states (DOS) (upper) and the resultant coupled electromagnetic ground states (lower). Here, FM, PM, and AFM indicate ferromagnetic, paramagnetic, and antiferromagnetic phases, respectively. (c-h) Theoretical calculation results of quasi-2D SRO as a function of in-plane strain within the DFT+DMFT framework at a temperature of 35 K. (c) Crystal field energy splitting energy ($h_{CF}$) between $d_{yz/zx}$ and $d_{xy}$ orbitals. (d) Occupancy of the $d_{yz/zx}$ and the $d_{xy}$ orbitals in the paramagnetic state. e) Orbital-resolved DOS at $E_F$ in the paramagnetic state. (f) Total DOS at $E_F$. (g) Contributions of orbitals ($d_{yz/zx}$ and $d_{xy}$) to ferromagnetic susceptibility ($\chi$). (h) Total $\chi$. See the main text for tendencies toward ferromagnetism (antiferromagnetism) by compressive (tensile) strains. Zero strain corresponds to the averaged in-plane lattice constant of bulk SRO and STO.

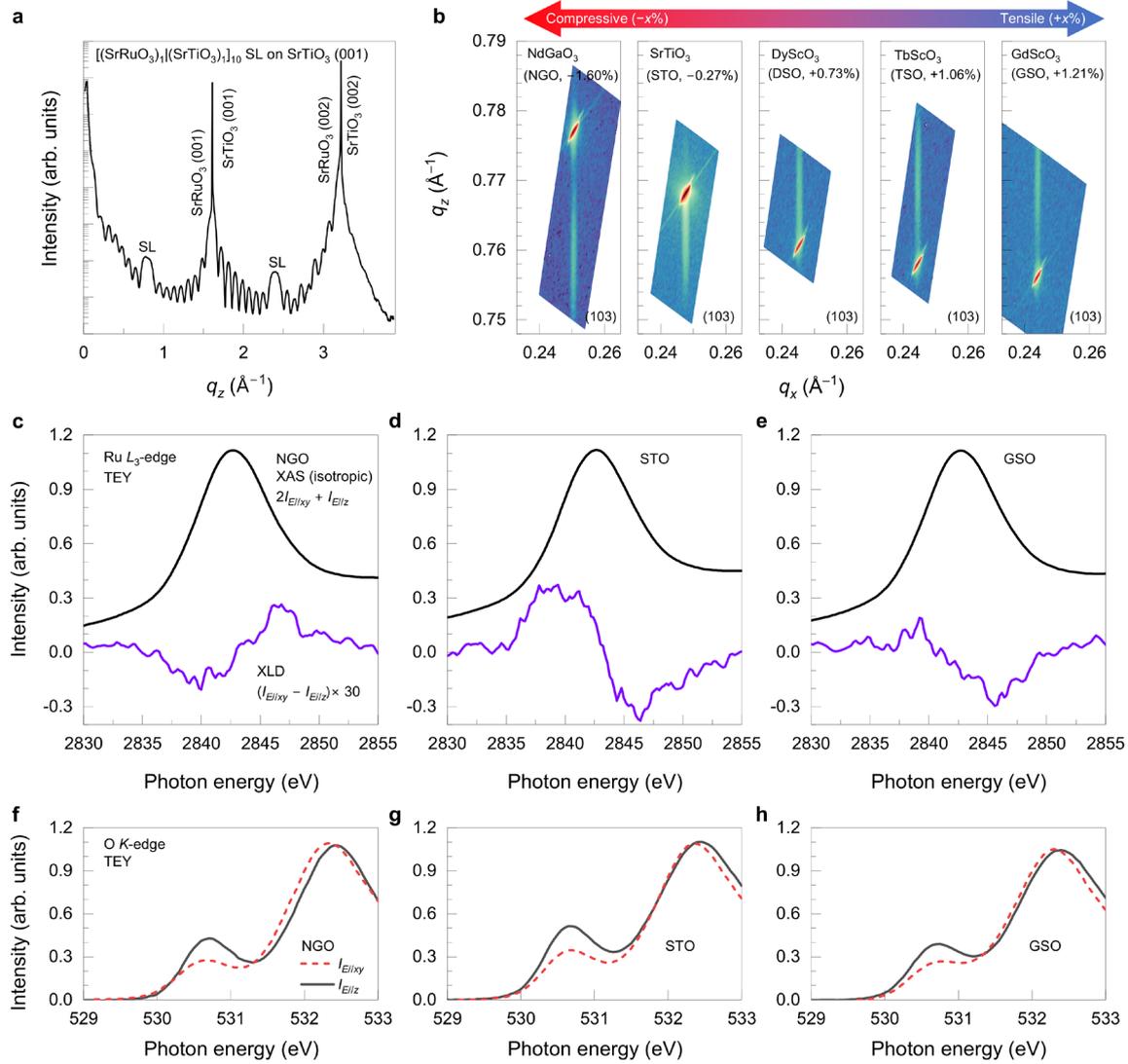

**Fig. 2. Structural characterization and orbital anisotropy modulation in strain-engineered quasi-2D SRO.** (a) Synchrotron XRD $\theta$–$2\theta$ scans of the atomically well-defined [(SRO)$_1$|(STO)$_1$]$_{10}$ superlattice on STO substrates. (b) XRD RSM results of quasi-2D superlattices with five different perovskite substrates, around the (103) Bragg reflection of each substrate, indicating the fully strained states. (c-e) Isotropic XAS and XLD spectra at Ru $L_3$-edge for quasi-2D SRO superlattices grown on (c) NGO, (d) STO, and (e) GSO substrates. (f-h) XAS spectra with two different polarizations at O $K$-edge for quasi-2D SRO superlattices grown on (f) NGO, (g) STO, and (h) GSO substrates. The O $K$-edge peaks at ~530.5 and ~532.5 eV primarily originate from Ru $4d$-O $2p$ and Ti $3d$-O $2p$ hybridization. XAS spectra are obtained by the total electron yield configuration.

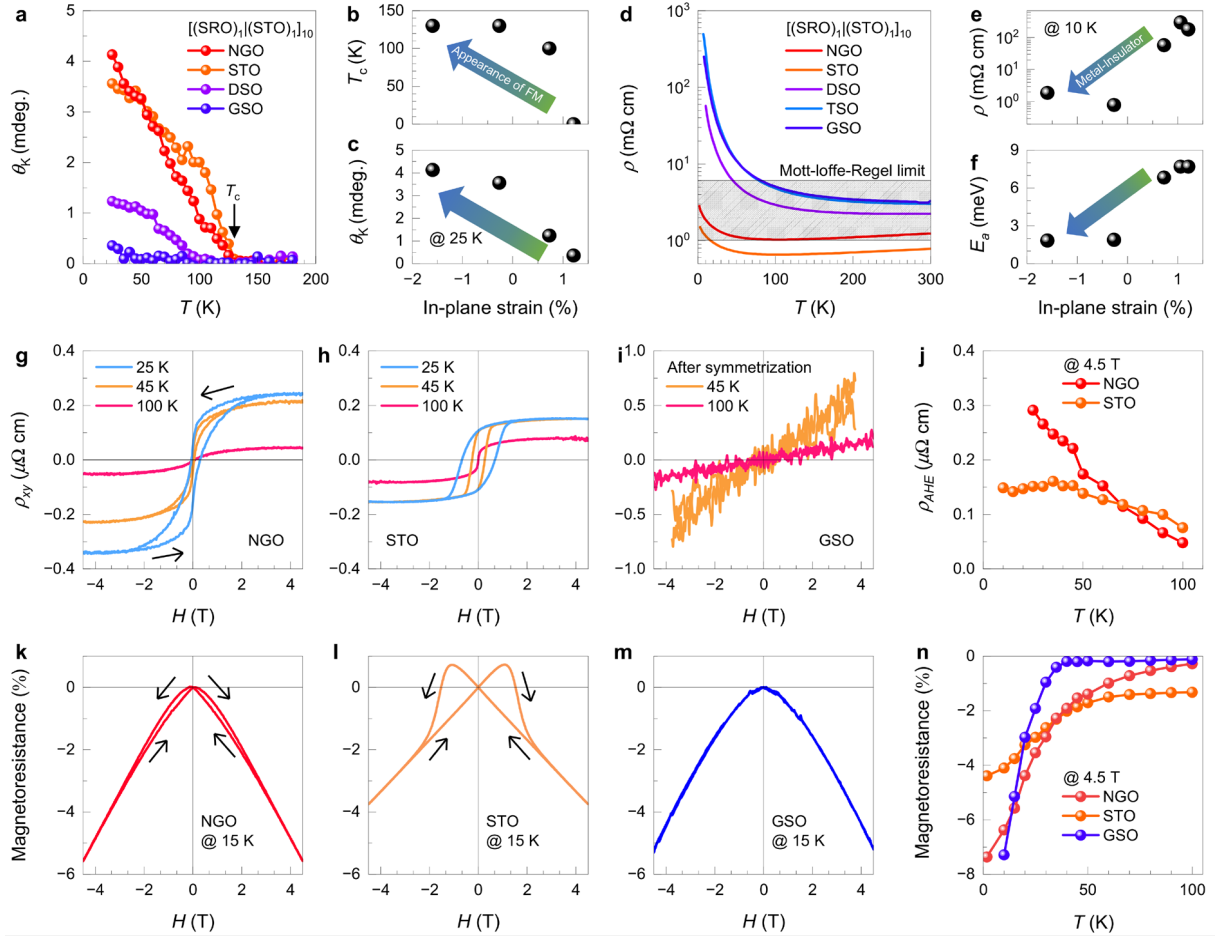

**Fig. 3. Strain-engineered vHs-induced electromagnetic phase transitions in quasi-2D SRO superlattices.** (a) $T$-dependent $\theta_K$ results and strain-dependent (b) $T_c$ and (c) $\theta_K$ of quasi-2D SRO superlattices obtained by polar MOKE measurements. (d-f) Metal-insulator transition in quasi-2D SRO superlattices appearing in (d) $T$-dependent $\rho$ and strain-dependent (e) $\rho$ at 10 K and (f) $E_a$. g-j $H$-dependent $\rho_{xy}$ of quasi-2D SRO superlattices grown on (g) NGO, (h) STO, and (i) GSO substrates at different $T$s, and (j) summary of $T$-dependent $\rho_{AHE}$ at $H$ = 4.5 T (see Fig. S13 for the extraction of $\rho_{AHE}$, supplementary material). For the GSO substrate case, due to the high $\rho$, we performed symmetrization. The data before symmetrization is shown in Fig. S12, supplementary material. (k-n) $H$-field dependent magnetoresistance of quasi-2D SRO superlattices grown on (k) NGO, (i) STO, and (m) GSO substrates at 15 K, and (n) summary of $T$-dependent magnetoresistance at $H$ = 4.5 T. The arrows indicate the direction of the $H$.